\documentclass[a4paper,fleqn]{cas-dc}

\usepackage{floatrow}

\usepackage{graphicx}
\usepackage[numbers,sort&compress]{natbib}

\usepackage[label font=bf,labelformat=simple]{subfig}

\def\tsc#1{\csdef{#1}{\textsc{\lowercase{#1}}\xspace}}
\tsc{WGM}
\tsc{QE}

\begin{document}
\let\WriteBookmarks\relax
\def\floatpagepagefraction{1}
\def\textpagefraction{.001}

\shorttitle{}    

\shortauthors{}  

\title [mode = title]{Control of free induction decay with quantum state preparation in a weakly coupled multi-spin system}  



%

\author[1]{Qian Cao}


\fnmark[1]




\affiliation[1]{organization={Key Laboratory of Artificial Micro- and Nano-structures of Ministry of Education}, 
            addressline={School of Physics and Technology, Wuhan University},
            city={Wuhan},
            state={Hubei},
            statesep={},
            postcode={430072}, 
            country={China} }

\author[1,2]{Tianzi Wang}
\affiliation[2]{organization={Hefei Natioanal Laboratory}, 
	city={Hefei},
      state={Anhui},
      statesep={},
	postcode={230088}, 
	country={China} }
\fnmark[1]




\author[1,3]{Wenxian Zhang}
\cormark[1]

\affiliation[3]{organization={Wuhan Institute of Quantum Technology}, 
            city={Wuhan},
            state={Hubei},
            statesep={},
            postcode={430206}, 
            country={China} }
\ead{wxzhang@whu.edu.cn}
\cortext[1]{Corresponding author}

\fntext[1]{These authors contribute equally.}


\begin{abstract}
Nuclear magnetic resonance (NMR) has been a widely used tool in various scientific fields and practical applications, with quantum control emerging as a promising strategy for synergistic advancements. In this paper, we propose a novel approach that combines NMR and quantum state preparation techniques to control free induction decay (FID) signals in weakly coupled spin systems, specifically Trifluoroiodoethylene $C_2F_3I$. We investigate the FID signal of the three-spin system and compare the differences between the FID signals in the thermal state and the pseudo-pure state (PPS), where the latter is generated using quantum state preparation techniques. Our approach aims to demonstrate a single exponentially decaying FID in weakly coupled spins, in which oscillatory FID signals are often observed. We validate our findings through numerical simulations and experimental measurements, and justify the validity of the theory. Our method opens a door to advancing spin system research and extending the capabilities of NMR with current quantum technologies in various scientific and practical fields.
\end{abstract}



\setlength{\mathindent}{0cm}
\maketitle

\section{Introduction}

NMR is a powerful technique that has played a crucial role in a wide range of scientific fields and practical applications. For instance, NMR has been pivotal in the development of quantum computing, magnetic resonance imaging, and quantum many-body theory in spin systems~\cite{callaghan1993principles,levitt2013spin,jones2000nmr,dale2015mri}.  Serving as an essential and adaptable instrument, NMR spectroscopy holds a broad array of applications, including the determination of molecular structures, monitoring of chemical reactions, and thus spurring advancements in drug discovery, materials science, and medical diagnostics~\cite{marion2013introduction,holzgrabe2005quantitative,clague1985review}. Over time, the development of numerous techniques related to NMR has led to substantial progress in these fields~\cite{ernst1987principles,keeler2011two,khaneja2005optimal}.

Over the past two decades, quantum control has progressively surfaced as a promising discipline, finding many applications in diverse areas such as quantum computation, quantum telecommunication, and quantum metrology~\cite{nielsen2000quantum,soare2014experimental}. Substantial progress has been achieved in various quantum systems including NV centers, ion traps, and superconducting circuits~\cite{dobrovitski2013quantum,monroe2013scaling,ma2021quantum}. The amalgamation of NMR and quantum control provides a potential pathway to develop more effective and innovative techniques across multiple fields~\cite{vandersypen2005nmr,ryan2008liquid}. This mutual interaction between NMR and quantum control might initiate breakthroughs in manipulating spin systems, thereby enriching their associated applications.The application of quantum control in NMR provides new opportunities for spin systems, thus, extends the capabilities of NMR techniques in various scientific and practical areas~\cite{schanda2016studying}.

As originally observed in experiments, FID reveals spin correlations, including the classical and the quantum correlations~\cite{Zobov2014, Liu2012}. Although a single exponential decay is expected in many NMR textbooks and papers~\cite{callaghan1993principles, Slichter1990, Chen2005}, oscillatory FID is often observed in real experiments, due to either long excitation-pulse effect or spin coupling in many-spin systems~\cite{Schenzle1980, Kunitomo1982, Boscaino1987, Sabirov1982, Cho2005}. With the aid of quantum control over the initial spin state, we investigate theoretically, numerically, and experimentally the FID signal from a thermal initial state and a PPS in a weakly coupled three-spin system. In multi-spin systems, the spin-spin coupling introduces challenges in discerning FID signals from those of an individual spin. However, in this paper, we advocate for the application of a quantum control method and the preparation of a unique initial quantum state to boost the FID signal in the many-spin system, specifically within $C_2F_3I$. Leveraging the power of quantum control techniques, our approach is meticulously designed to improve the quality and dependability of FID signals, subsequently allowing for a deeper exploration and understanding of coupled spin systems. Considering the significant usage of quantum control in spectroscopy \cite{oron2002quantum,wollenhaupt2005quantum,dudovich2004quantum}, we anticipate the potential extension of this technique beyond the scope of NMR spectrum to further areas. 

The paper is organized as follows: Sec.~\ref{sec:h} details the Hamiltonian and parameters of the three-spin system employed in our experiments. The FID of a single spin is subsequently reviewed in Sec.~\ref{sec:FID}. In Sec.~\ref{sec:therm} and~\ref{sec:pps}, we present respectively theories and numerical simulations to show the differences between the FID signals arising from the thermal state (TS) and a pseudo-pure state (PPS), where the latter leverages quantum control to prepare the quantum state. Our results are validated further through experimental measurements in Sec.~\ref{sec:exp}. Finally, Sec.~\ref{sec:con} summarizes our results and discusses the valid coupling strength of our theory.

\section{Spin Hamiltonian}
\label{sec:h}

Consider a system with three fluorine nuclear spins of a $C_2F_3I$ liquid in a Tesla magnetic field. The spin Hamiltonian is (setting $\hbar = 1$) 
\begin{equation}
H=\sum_{i=1}^3(\omega_0+\delta_i)I_{iz}+\sum_{i<j}J_{ij}{\bf I_i}\cdot {\bf I_j}+\sum_{i=1}^3{\bf \eta}\cdot {\bf I_i}
\label{eq:H}
\end{equation}
where $\omega_0 = \gamma B_0$ represents the angular Larmor frequency in a strong bias magnetic field $B_0$ with $\gamma$ the nuclear gyromagnetic ratio and $\delta_i$ represents the chemical shifts of each spin. The coefficient $J_{ij}$ is the J-coupling strength in an isotropic form between each pair of spins ${\bf I_i}$ and ${\bf I_j}$~\cite{levitt2013spin,abragam1961principles}, and $\bf \eta$ is the random magnetic field noise. In our NMR experiment, we set $\delta_1=0$ and find $\delta_2=-1393$ Hz, $\delta_3=1027$ Hz, and $J_{12}=-130$ Hz, $J_{13}=69$ Hz, and $J_{23}=50$ Hz.

In a rotating reference frame defined by $e^{-i\omega_0 t\sum_i I_{iz}}$, the Zeeman term with $\omega_0$ is canceled. Since the parameters of the system lie in the weak coupling regime, $|\delta_i-\delta_j| \gg |J_{ij}|$, the transversal coupling terms $I_{ix}I_{jx}+I_{iy}I_{jy}$ oscillate rapidly and average to zero. Furthermore, $\omega_0 \gg |{\bf\eta}|$ in the experiment, the secular term $\eta_{z}I_{iz}$ (with $\eta_z \sim 20$ Hz) dominates, and the effects of $xy$ components are small enough to be safely neglected. Therefore, the effective Hamiltonian in the rotating reference frame becomes
\begin{equation}
H_E \approx \sum_{i=1}^3\delta_i I_{iz}+\sum_{i<j}J_{ij} I_{iz} I_{jz}+\sum_{i=1}^3 \eta_{z} I_{iz}.
\label{eq:HE}
\end{equation}

\section{Review on FID of an ideal spin}
\label{sec:FID}

We begin by considering the FID of a single spin with pure dephasing before discussing the multi-spin system. In the rotating reference frame, the Hamiltonian is
\begin{equation}
H_E \approx \eta_z I_z.
\end{equation}
The longitudinal noise $\eta_z$ randomly modulates the spin precession frequency along the $z$ axis and induces pure dephasing \cite{yang2016quantum}, where $\eta_z$ is assumed a static noise.

The density matrix of a thermal equilibrium state is 
\begin{equation}
\rho_T=\frac{e^{-H/k_BT}}{Tr(e^{-H/k_BT})} \approx \frac{1}{2}(I+p\sigma_{z})
\label{eq:rhoT}
\end{equation}
where $H = (\omega_0 + \eta_z) I_z $ is the spin Hamiltonian in laboratory reference frame, $p=-{\omega_0}/{(2k_BT)}$ and $\sigma_z=2I_z$ is the Pauli matrix. The above approximation holds under the condition that at room temperature, $k_BT \gg \hbar\omega_0$. After a $\pi/2$ rotation along the $y$ axis, the initial density matrix becomes $\frac{1}{2}(I+p\sigma_{x})$.

Traditionally, in NMR experiments, the FID signal is proportional to the transverse magnetization, which can be expressed as a standard quantum-mechanical average~\cite{messiah1961quantum}. For an ensemble of nuclear spins~\cite{dobrovitski2008decoherence}, this average needs to be taken with static random noise $\eta_z$. The $x$ and $y$ components of the transverse magnetization represent the real part and imaginary parts of FID signal respectively, which are given by
\begin{equation}
\begin{split}
&M_x=\left \langle Tr(U\rho_0 U^\dagger \sigma_{x}) \right \rangle =\frac{p}{2}\,\overline{\cos \eta_z t}, \\
&M_y=\left \langle Tr(U\rho_0 U^\dagger \sigma_{y}) \right \rangle =-\frac{p}{2}\,\overline{\sin \eta_z t}
\label{eq:Mxy}
\end{split}
\end{equation}
where $U=\exp(-iH_{E}t)$ and $M_x(t)$ and $M_y(t)$ are the averages over $\eta_z$ of $\cos(\eta_z t)$ and $\sin(\eta_z t)$, respectively. Here, $\overline{(\cdot)}$ denotes the statistical ensemble average over many nuclear spins. The amplitude of the transverse magnetization, also the modulus length of FID signal, is
\begin{equation}
M_{\perp}=-\frac{p}{2}\sqrt{\left(\overline{\cos \eta_z t}\right)^2+\left(\overline{\sin \eta_z t}\right)^2}\; .
\label{eq:Mp}
\end{equation}

The amplitude of transverse magnetization, as denoted by Eq. (6), evolves under the random noise $\eta_z$ present in the ensemble of individual spins, assuming that spin interactions are negligible. Various types of classical noise exhibit distinct forms. For instance, white noise, Gaussian noise, and Lorentzian noise has a probability density function (with zero average) as follows,

\begin{equation}
P(\eta_z)=\left\{
\begin{aligned}
&\frac{1}{2a} , \quad |\eta_z|\le a, \\
&\frac{1}{\sqrt{2\pi}}\exp(-\frac{\eta_z^2}{2\sigma^2}), \\
&\frac{1}{\pi}\frac{\gamma}{\eta_z^2+\gamma^2}.
\end{aligned}
\right.
\end{equation}

\begin{figure}
	\centering
	\includegraphics[width=3.2 in, height=6cm]{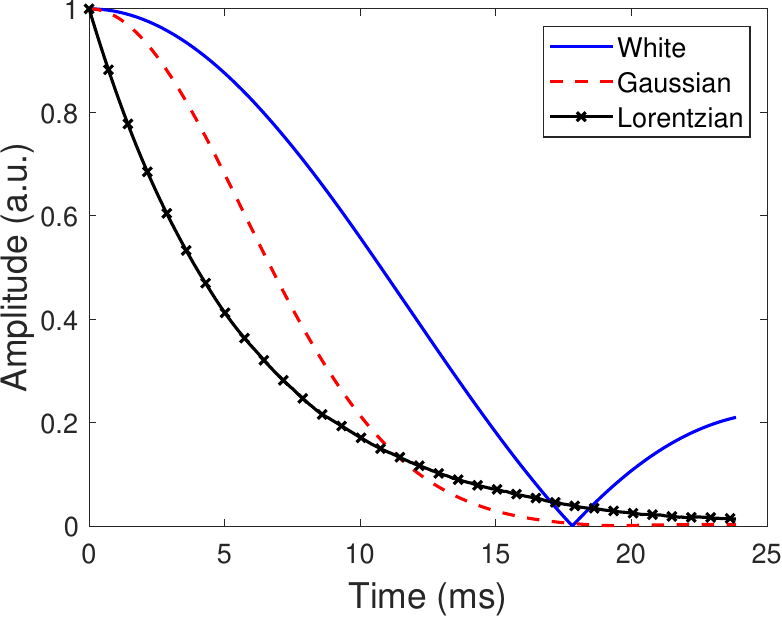}
	\caption{FID of a single spin subject to different kind of noises. The blue solid line, red dashed line, and black solid line with crosses correspond to classical white noise ($a=28$ Hz), Gaussian noise ($\sigma=28$ Hz), and Lorentzian noise ($\gamma=28$ Hz), respectively.}
	\label{fig:1}
\end{figure}

In Fig.~\ref{fig:1}, we simulate the evolution of the FID amplitude for these three types of classical noise. The FID for white noise resembles the absolute value of a sinc function, while that of Gaussian noise maintains a Gaussian profile. The FID for Lorentzian noise exhibits an exponential decay, identical to the result after Fourier transformation. The variations in noise distribution contribute to differences in FID signals. In our experiment, the most closely resembling simulated noise is Lorentzian noise. When considering the ensemble of individual spins, the situation is simple, as only noise produces effects. However, the conditions change when considering interacting spins in a molecular ensemble.

\section{FID of thermal state of $C_2F_3I$}
\label{sec:therm}

For the $C_2F_3I$ spin system discussed in Eq.~(\ref{eq:H}), the density matrix of the thermal state is an extension of Eq.~(\ref{eq:rhoT}) \cite{cory2000nmr}. Due to the fact that $H/k_BT$ is of the order of $10^{-5}$, the NMR spin system is in a fully mixed state at room temperature. With the approximation that $\omega_0 \gg |\delta_i| \gg |J_{ij}|$, the coupling between spins can be neglected, thus
\begin{equation}
\rho_T\approx \frac{I}{8}+\frac{p} 8 (\sigma_{1z}+\sigma_{2z}+\sigma_{3x})
\end{equation}
where $\rho_T$ represents the thermal initial state (TS) after a $\pi/2$ rotation on spin $F_3$ along the $y$ axis. Under the effective Hamiltonian $H_E$ Eq.~(\ref{eq:HE}), one finds that spins $F_1$ and $F_{2}$ do not contribute to the FID signal and the transverse magnetization can be calculated using a similar formula as Eq.~(\ref{eq:Mxy}).

The transverse magnetization is obtained
\begin{equation}
\begin{split}
   &M_x=-M_0(t) \left(\cos(\delta_3t)\overline{\cos \eta_{z} t}-\sin(\delta_3t)\overline{\sin \eta_{z} t}\right), \\
   &M_y=M_0(t) \left(\cos(\delta_3t)\overline{\sin \eta_{z} t}+\sin(\delta_3t)\overline{\cos \eta_{z} t}\right)
\end{split}   
\end{equation}
where $M_0(t) = -(p/2)\cos({J_{13}t}/2)\cos({J_{23}t}/2)$. The amplitude of the FID signal is 
\begin{equation}
\label{eq:ts}
M_{\perp}=|M_0(t)| \sqrt{\left(\overline{\cos \eta_{z} t}\right)^2+\left(\overline{\sin \eta_{z} t}\right)^2}\;.
\end{equation}
In comparison to Eq.~(\ref{eq:Mp}), which neglects $J_{13}$ and $J_{23}$, the FID signal is nearly identical. However, when $J_{13}$ and $J_{23}$ cannot be neglected, the FID signal deviates from the standard FID signal and is disrupted by two cosine terms of spin interaction.

\begin{figure}
\centering
\includegraphics[height = 6cm, width = 3.2 in]{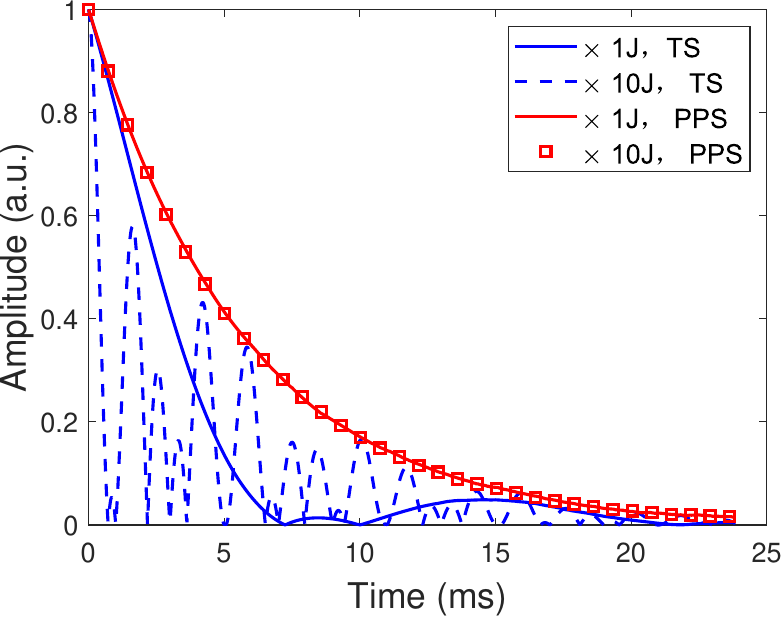}
\caption{FID in the thermal state (TS) and the PPS with different coupling strength. The noise is assumed Lorentzian line shape noise with $\gamma=28$Hz. The red solid line with squares and the blue solid line respectively represent the FID signals in TS and PPS with experimental parameters in Sec.~\ref{sec:h}, while the red squares denote the FID signal with $J_{ij}$  magnified by a factor of 10 in the TS. The blue dashed line represents the FID signal in the PPS $|101\rangle$ with $J_{ij}$ magnified by a factor of 10. The numerical results are averaged over $10^5$ random realizations.}
\label{fig:2}
\end{figure}

The FID signals of a thermal initial state are simulated with the effective Hamiltonian Eq.~(\ref{eq:HE}) and plotted in Fig.~\ref{fig:2} for different coupling strength. Compared to the ideal curve shown in Fig.~\ref{fig:1}, one immediately finds the modulation effect of $M_0(t)$. Furthermore, the larger the coupling strength, the rapider the modulation.

\section{FID of Pseudo-pure state (PPS) of $C_2F_3I$}
\label{sec:pps}

In the field of NMR, it is desirable to prepare a pure state of an ensemble for experimental investigations. However, at room temperature, it is difficult to generate a pure state due to thermal relaxation, leading to the use of PPS in most NMR quantum information processing experiments~\cite{jones2010quantum}. PPS is such a mixed state that differs from a pure state by only an identity matrix, therefore both states show exactly the same FID signal since the identity matrix contributes nothing to the FID.

To prepare a PPS, the spatial or temporal average method is typically employed~\cite{cory2000nmr, cory1998nuclear}. For example, a PPS of $C_2F_3I$ in our experiments is prepared by using the temporal average method
\begin{equation}
 	\rho_{PPS}=\frac{1-p}{8}I+p|101\rangle\langle101|,
\end{equation}
which has the same observation effect as a pure state $|101\rangle\langle101|$ due to the identity operator having no effect on Pauli matrices. Here the spin state $|0\rangle$ ($|1\rangle$) is the eigenstate of $\sigma_z$ with the eigenvalue $+1$ ($-1$).

After preparing the PPS state, a $\pi/2$ rotation on spin $F_3$ along the $y$ axis produces the FID as
\begin{equation}
\begin{split}
	&M_x=p\left(\cos\theta \, \overline{\cos \eta_{z} t}-\sin\theta \, \overline{\sin \eta_{z} t}\right),\\
	&M_y=-p\left(\sin\theta \, \overline{\cos \eta_{z} t}+\cos\theta \, \overline{\sin \eta_{z} t}\right)
\end{split}
\end{equation}
where $\theta(t)=(\delta_3-J_{13}/2+J_{23}/2)t$. The amplitude of FID signal becomes
\begin{equation}
\label{eq:pps}
	M_{\perp}=-p\sqrt{\left(\overline{\cos \eta_{z} t}\right)^2+\left(\overline{\sin \eta_{z} t}\right)^2}\;.
\end{equation}

The amplitude of the FID signal is exactly the same as the ideal FID signal (see Eq.~(\ref{eq:Mp}) and in Fig.~\ref{fig:2}), suggesting that the PPS of a multiple interacting spin system after quantum state preparation can discard the disturbance of $J$ coupling between spins and chemical shifts, no matter how large they are once the effective Hamiltonian $H_E$ is valid. Therefore, we propose an easy method to observe the ideal FID curve of a coupled spin system by preparing the spin system in an initial PPS. Of course, to validate the effective Hamiltonian, one needs to satisfy the approximation $\omega_0 \gg |\delta_i-\delta_j| \gg |J_{ij}|$ and $\omega_0 \gg |\eta_z|$.

\begin{figure}
	\centering
	\includegraphics[width=8 cm, height=6cm]{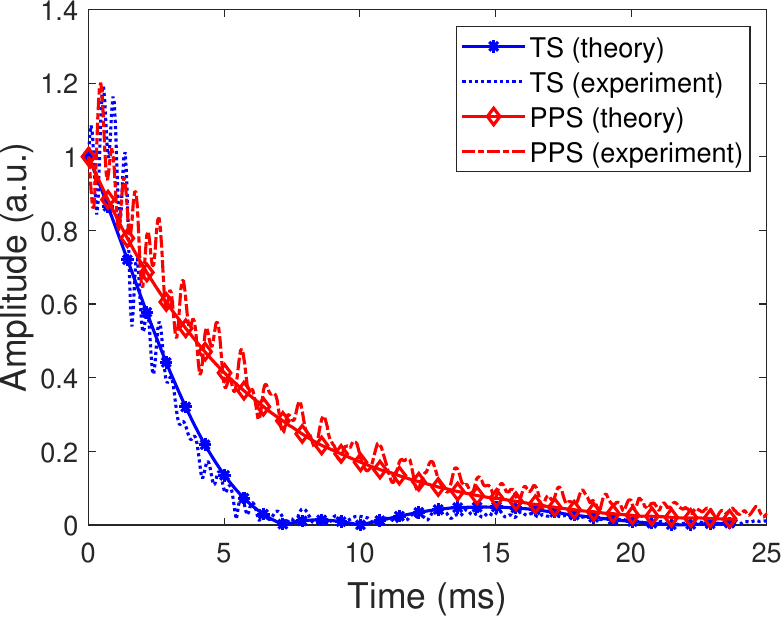}
	\caption{FID in experiment and theory. The blue line with asterisks and red line with diamonds depict the FID signal for the thermal and PPS state ($|101\rangle$), respectively, under Lorentzian noise ($\gamma=28$ Hz). The blue dotted and red dashed lines correspond to the experimental results of $C_2F_3I$, averaged over 100 runs.}
	\label{fig:exp}
\end{figure}

\section{Experimental results}
\label{sec:exp}

We carry out experimental investigations on the FID of three fluorine nuclear spins in the molecule $C_2F_3I$. The relaxation time for the spins is in the order of $10$ s~\cite{Luo16}. To prepare the thermal initial state, we wait for a long time (180 s) to allow the spins to relax completely, then apply a selective $\pi/2$ pulse along $y$ direction on the spin $F_3$ with a pulse duration of 3 ms and acquire the FID signal. We average the FIDs over 100 runs of experiments to reduce the statistical error. For the PPS experiments, the procedure is the same, except that preparing the PPS with temporal averaging method~\cite{Knill97}, before applying the $\pi/2$ pulse on the spin $F_3$. Figure~\ref{fig:exp} illustrates the experimental results from the thermal state and the PPS. To compare, the theoretical predictions are also plotted, Eq.~(\ref{eq:ts}) for the thermal state and Eq.~(\ref{eq:pps}) for the PPS with the Lorentzian noise model. The decay rate $\gamma = 28$ Hz is adjusted manually to fit the experimental results.

As shown in Fig.~\ref{fig:exp}, the theoretical predictions agree overall with the experimental results for both the thermal state and the PPS. The exponential decay of experimental results for both initial states and the modulation of spin coupling for the thermal state are captured by the weak coupling theory. However, different from theoretical smooth lines, the experimental results show many regular, small and rapid oscillations. The discrepancy might come from the violation of the weak coupling approximation. As shown by our analysis and numerical simulations in Append.~\ref{sec:appd}, the Heisenberg form spin coupling $J_{ij} {\bf I_i}\cdot {\bf I_j}$ indeed introduces oscillations with frequencies in the order of $|\delta_i-\delta_j|$, however, the oscillation amplitudes are much smaller than those observed in experiments. Thus, the physical mechanism behind these experimental oscillations still needs to be explored.

\section{Conclusions and discussions}
\label{sec:con}

In summary, we demonstrate the improved FID signal via quantum state preparation technique in a weakly coupled three-spin system. Theoretical analysis shows that in a coupled spin system, J-coupling terms modulate the FID signal, causing it to deviate from the single exponential decay expected in ideal spin systems. By preparing the spin system in a designed PPS with quantum control, the ideal FID signal is recovered, with the interference caused by the $J$-coupling terms canceled, and becomes independent of the spin coupling strength and the bias magnetic field. Results from numerical simulations and NMR experiments in liquid $C_2F_3I$ agree with theoretical predictions. This leads to a better understanding of the system dynamics and enhances the accuracy of information extracted from the FID signal.

We note that the bias magnetic field in our experiments is rather low, 1 T, compared to other high field experiments with $10$ T~\cite{Luo16}. However, higher bias magnetic field increases the difference of chemical shift $|\delta_i-\delta_j|$, and consequently the ratio of the coupling strength and chemical shift difference $J_{ij}/|\delta_i-\delta_j|$ is smaller, thus one can safely expect that the theory under the weak coupling condition holds even better.

The implementation of quantum state preparation in weakly coupled multi-spin systems provides promising avenues for enhancing the efficiency of various NMR techniques, encompassing quantum computing, MRI, and the investigation of spin system properties. Potential future work could extend this approach to other coupled spin systems, optimize the PPS preparation process, increase the bias magnetic field, and explore the capabilities of more advanced quantum control methods in NMR to further amplify signal quality across other spectroscopy domains.

\section*{Acknowledgements}
The work is supported by National Natural Science Foundation of China (NSFC) under Grant No. 12274331,  the NSAF under Grant No. U1930201, and Innovation Program for Quantum Science and Technology under Grant No. 2021ZD0302100. The numerical calculation in the paper have been partially done on the supercomputing system in the Supercomputing Center of Wuhan University.








\appendix

\section{Beyond the weak coupling regime}
\label{sec:appd}

\begin{figure}
\centering
\sidesubfloat{\includegraphics[height = 6cm, width = 8 cm]{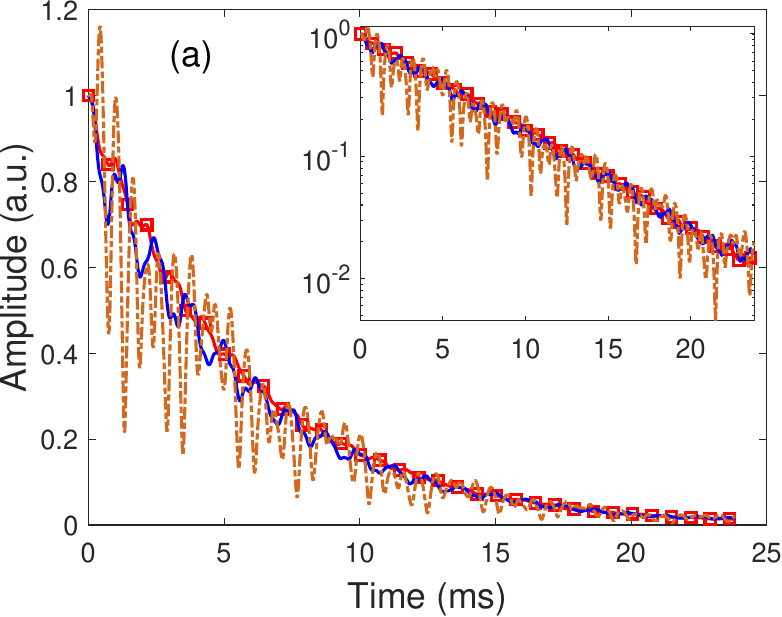}}\\
\sidesubfloat{\includegraphics[height = 6cm, width = 8 cm]{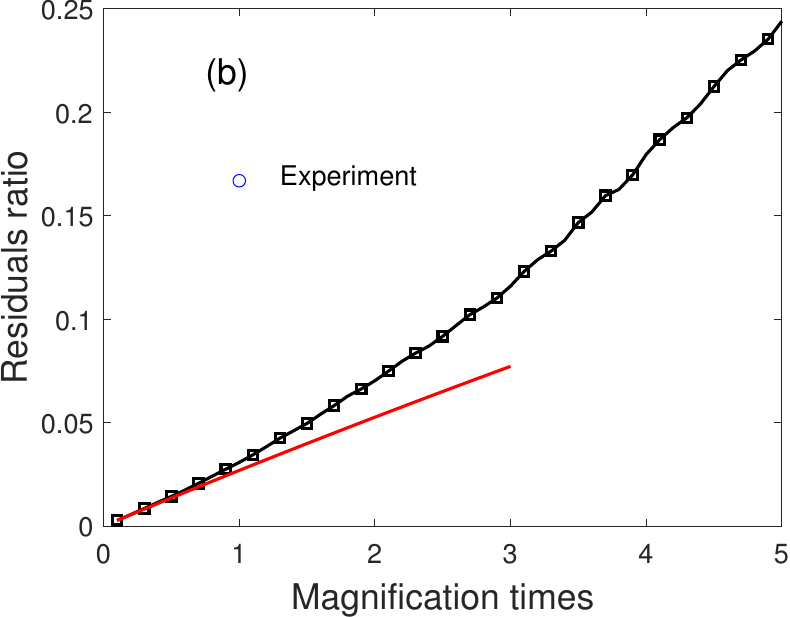}}
\caption{Numerical simulation of FID beyond the weak coupling regime. (a) FID with a magnification factor of 1 (red solid line with squares), 2.5 (blue solid line) and 5 (brown dash-doted line) on the J coupling under the Lorentzian noise ($\gamma=28$ Hz). The inset shows the same result in logarithm scale. (b) Residual ratio as a function of the magnification times $m$ of spin coupling strength from numerical simulations (solid line with squares) and prediction from Eq.~(\ref{eq:rr}) (red solid line). The numerical results are averaged over $10^5$ statistical runs. Experimental result calculated from Fig.~\ref{fig:exp} is shown as a blue circle. The evolution time $T$ is 24 ms.}
\label{fig:bey}
\end{figure}

Weak coupling criterion is usually applicable to heteronuclear spins and small homonuclear molecules~\cite{vandersypen2005nmr}. However, the experimental results and theoretical predictions under the weak coupling condition, as shown in Fig.~\ref{fig:exp}, do not agree with each other, especially in the rapid oscillations in experimental FID. Therefore, we explore instances where the coupling constant is not minor, resulting in a Hamiltonian in the following form in a rotating reference frame defined by $\exp(-it\omega_0\sum_i I_{iz})$

\begin{equation}
\label{eq:HR}
	H_R = \sum_{i=1}^3\delta_iI_{iz}+m\sum_{i<j}J_{ij} {\bf I_i}\cdot {\bf I_j}+\sum_{i=1}^3 \eta_{z}\, I_{iz},
\end{equation}
where the magnification times $m$ effectively enlarges the coupling constant $J_{ij}$.

Beyond the weak coupling regime, the spin system described by the above Hamiltonian Eq.~(\ref{eq:HR}) is not analytical solvable. We thus employ numerical method to simulate the spin system and calculate the FID. We focus on the initial PPS. All parameters and noises are the same as in Fig.~\ref{fig:exp} except that the coupling form is changed. 

The numerical results are shown in Fig.~\ref{fig:bey}(a). Obviously, for small $m=1$, one barely see oscillations around the exponential decay. However, as $m$ becomes large, the oscillations around the decay are clearly observed. Moreover, the oscillation frequency is in the order of $|\delta_i-\delta_j| \sim 10^3$ Hz.

To quantify the deviation of the simulated FID from the ideal FID which corresponds to $m=0$, we define residual ratio
\begin{equation}
\label{eq:rr}
R(m) = \frac{\int_0^T |A_m(t)-A_0(t)| dt}{\int_0^T A_0(t) dt}
\end{equation}
where $T$ is the total evolution time and $A_m$ is the FID. The results of residual ratio are plotted in Fig.~\ref{fig:bey}(b) and shows a monotonic increase as $m$ becomes large.

We employ time-independent perturbation theory and separate the Hamiltonian Eq.~(\ref{eq:HR}) as $H_R=H_E+H_1$, where $H_E$ represents the Hamiltonian in Eq.~(\ref{eq:HE}) but with an $m$ (not necessarily $m=1$) and 
\begin{equation*}
H_1=m\sum_{i<j} J_{ij}(I_{ix}I_{jx}+I_{iy}I_{jy})
\end{equation*}
is regarded as the perturbation term. We account for the first-order perturbation and retain the first order of small quantity $mJ_{ij}/|\delta_i-\delta_j|$, culminating in the result
\begin{equation}
	\begin{split}
		&M_x = -p[(1-\lambda_1-\lambda_2) \cos \theta_3 +\lambda_1 \cos \theta_2+
		\lambda_2 \cos \theta_1], \\
		&M_y = p[(1-\lambda_1-\lambda_2) \sin \theta_3 +\lambda_1 \sin \theta_2+
		\lambda_2 \sin \theta_1]
	\end{split}
\end{equation}
where $\theta_1 = [\delta_1+\eta_z + m(J_{12}-J_{13})/2]t$, $\theta_2 = [\delta_2+\eta_z + m(J_{23}-J_{12})/2]t$, and $\theta_3 = [\delta_3+\eta_z+m(J_{23}-J_{13})/2]t$. Here, $\lambda_1=mJ_{23}/[2(\delta_3-\delta_2)]$ and $\lambda_2 = -mJ_{13}/[2(\delta_3-\delta_1)]$. 

The FID signal can be reexpressed as
\begin{equation}
	M_{\perp} = -p F(J_{ij},\delta_i) \times \sqrt{\left(\overline{\cos \eta_z t}\right)^2+\left(\overline{\sin \eta_z t}\right)^2},
\end{equation}
where
\begin{equation}
	F(J_{ij}, \delta_i) \approx 1-\lambda_1-\lambda_2 +\lambda_1 \cos((\delta_3-\delta_2)t) +\lambda_2 \cos(\delta_3t).
\end{equation}
We have adopt that $\delta_1 = 0$. Therefore one finds immediately that $R \propto m$ as illustrated in Fig.~\ref{fig:bey}(b).

    

\printcredits

\bibliographystyle{model1-num-names}
\bibliography{fid}

\begin{thebibliography}{39}
\expandafter\ifx\csname natexlab\endcsname\relax\def\natexlab#1{#1}\fi
\providecommand{\url}[1]{\texttt{#1}}
\providecommand{\href}[2]{#2}
\providecommand{\path}[1]{#1}
\providecommand{\DOIprefix}{doi:}
\providecommand{\ArXivprefix}{arXiv:}
\providecommand{\URLprefix}{URL: }
\providecommand{\Pubmedprefix}{pmid:}
\providecommand{\doi}[1]{\href{http://dx.doi.org/#1}{\path{#1}}}
\providecommand{\Pubmed}[1]{\href{pmid:#1}{\path{#1}}}
\providecommand{\bibinfo}[2]{#2}
\ifx\xfnm\relax \def\xfnm[#1]{\unskip,\space#1}\fi
\bibitem[{Callaghan(1993)}]{callaghan1993principles}
\bibinfo{author}{P.~T. Callaghan}, \bibinfo{title}{Principles of Nuclear
  Magnetic Resonance Microscopy}, \bibinfo{publisher}{Oxford University Press},
  \bibinfo{year}{1993}.
\bibitem[{Levitt(2013)}]{levitt2013spin}
\bibinfo{author}{M.~H. Levitt}, \bibinfo{title}{Spin Dynamics: Basics of
  Nuclear Magnetic Resonance}, \bibinfo{publisher}{John Wiley \& Sons},
  \bibinfo{year}{2013}.
\bibitem[{Jones(2000)}]{jones2000nmr}
\bibinfo{author}{J.~A. Jones},
\newblock \bibinfo{title}{{NMR} quantum computation},
\newblock \bibinfo{journal}{arXiv preprint quant-ph/0009002}
  (\bibinfo{year}{2000}).
\bibitem[{Dale et~al.(2015)Dale, Brown, and Semelka}]{dale2015mri}
\bibinfo{author}{B.~M. Dale}, \bibinfo{author}{M.~A. Brown},
  \bibinfo{author}{R.~C. Semelka}, \bibinfo{title}{MRI: Basic Principles and
  Applications}, \bibinfo{publisher}{John Wiley \& Sons}, \bibinfo{year}{2015}.
\bibitem[{Marion(2013)}]{marion2013introduction}
\bibinfo{author}{D.~Marion},
\newblock \bibinfo{title}{An introduction to biological nmr spectroscopy},
\newblock \bibinfo{journal}{Mol. Cell. Proteomics} \bibinfo{volume}{12}
  (\bibinfo{year}{2013}) \bibinfo{pages}{3006--3025}.
\bibitem[{Holzgrabe et~al.(2005)Holzgrabe, Deubner, Schollmayer, and
  Waibel}]{holzgrabe2005quantitative}
\bibinfo{author}{U.~Holzgrabe}, \bibinfo{author}{R.~Deubner},
  \bibinfo{author}{C.~Schollmayer}, \bibinfo{author}{B.~Waibel},
\newblock \bibinfo{title}{Quantitative {NMR} spectroscopy—applications in
  drug analysis},
\newblock \bibinfo{journal}{J. Pharm. Biomed. Anal.} \bibinfo{volume}{38}
  (\bibinfo{year}{2005}) \bibinfo{pages}{806--812}.
\bibitem[{Clague(1985)}]{clague1985review}
\bibinfo{author}{A.~D. Clague},
\newblock \bibinfo{title}{Review of the applications of recent developments of
  {NMR} in materials science},
\newblock \bibinfo{journal}{Helv. Phys. Acta;(Switzerland)}
  \bibinfo{volume}{58} (\bibinfo{year}{1985}).
\bibitem[{Ernst et~al.(1987)Ernst, Bodenhausen, and
  Wokaun}]{ernst1987principles}
\bibinfo{author}{R.~Ernst}, \bibinfo{author}{G.~Bodenhausen},
  \bibinfo{author}{A.~Wokaun}, \bibinfo{title}{Principles of Nuclear Magnetic
  Resonance in One and Two Dimensions}, International series of monographs on
  chemistry, \bibinfo{publisher}{Oxford University Press},
  \bibinfo{year}{1987}.
\bibitem[{Keeler(2011)}]{keeler2011two}
\bibinfo{author}{J.~Keeler}, \bibinfo{title}{Understanding NMR Spectroscopy},
  \bibinfo{publisher}{John Wiley \& Sons}, \bibinfo{year}{2011}.
\bibitem[{Khaneja et~al.(2005)Khaneja, Reiss, Kehlet, Schulte-Herbr{\"u}ggen,
  and Glaser}]{khaneja2005optimal}
\bibinfo{author}{N.~Khaneja}, \bibinfo{author}{T.~Reiss},
  \bibinfo{author}{C.~Kehlet}, \bibinfo{author}{T.~Schulte-Herbr{\"u}ggen},
  \bibinfo{author}{S.~J. Glaser},
\newblock \bibinfo{title}{Optimal control of coupled spin dynamics: design of
  {NMR} pulse sequences by gradient ascent algorithms},
\newblock \bibinfo{journal}{J. Magn. Reson.} \bibinfo{volume}{172}
  (\bibinfo{year}{2005}) \bibinfo{pages}{296--305}.
\bibitem[{Nielsen and Chuang(2000)}]{nielsen2000quantum}
\bibinfo{author}{M.~Nielsen}, \bibinfo{author}{I.~Chuang},
  \bibinfo{title}{Quantum Computation and Quantum Information},
  \bibinfo{publisher}{Cambridge University Press}, \bibinfo{year}{2000}.
\bibitem[{Soare et~al.(2014)Soare, Ball, Hayes, Sastrawan, Jarratt, McLoughlin,
  Zhen, Green, and Biercuk}]{soare2014experimental}
\bibinfo{author}{A.~Soare}, \bibinfo{author}{H.~Ball},
  \bibinfo{author}{D.~Hayes}, \bibinfo{author}{J.~Sastrawan},
  \bibinfo{author}{M.~Jarratt}, \bibinfo{author}{J.~McLoughlin},
  \bibinfo{author}{X.~Zhen}, \bibinfo{author}{T.~Green},
  \bibinfo{author}{M.~Biercuk},
\newblock \bibinfo{title}{Experimental noise filtering by quantum control},
\newblock \bibinfo{journal}{Nat. Phys.} \bibinfo{volume}{10}
  (\bibinfo{year}{2014}) \bibinfo{pages}{825--829}.
\bibitem[{Dobrovitski et~al.(2013)Dobrovitski, Fuchs, Falk, Santori, and
  Awschalom}]{dobrovitski2013quantum}
\bibinfo{author}{V.~Dobrovitski}, \bibinfo{author}{G.~Fuchs},
  \bibinfo{author}{A.~Falk}, \bibinfo{author}{C.~Santori},
  \bibinfo{author}{D.~Awschalom},
\newblock \bibinfo{title}{Quantum control over single spins in diamond},
\newblock \bibinfo{journal}{Annu. Rev. Condens. Matter Phys.}
  \bibinfo{volume}{4} (\bibinfo{year}{2013}) \bibinfo{pages}{23--50}.
\bibitem[{Monroe and Kim(2013)}]{monroe2013scaling}
\bibinfo{author}{C.~Monroe}, \bibinfo{author}{J.~Kim},
\newblock \bibinfo{title}{Scaling the ion trap quantum processor},
\newblock \bibinfo{journal}{Science} \bibinfo{volume}{339}
  (\bibinfo{year}{2013}) \bibinfo{pages}{1164--1169}.
\bibitem[{Ma et~al.(2021)Ma, Puri, Schoelkopf, Devoret, Girvin, and
  Jiang}]{ma2021quantum}
\bibinfo{author}{W.-L. Ma}, \bibinfo{author}{S.~Puri}, \bibinfo{author}{R.~J.
  Schoelkopf}, \bibinfo{author}{M.~H. Devoret}, \bibinfo{author}{S.~M. Girvin},
  \bibinfo{author}{L.~Jiang},
\newblock \bibinfo{title}{Quantum control of bosonic modes with superconducting
  circuits},
\newblock \bibinfo{journal}{Sci. Bull.} \bibinfo{volume}{66}
  (\bibinfo{year}{2021}) \bibinfo{pages}{1789--1805}.
\bibitem[{Vandersypen and Chuang(2005)}]{vandersypen2005nmr}
\bibinfo{author}{L.~M. Vandersypen}, \bibinfo{author}{I.~L. Chuang},
\newblock \bibinfo{title}{{NMR} techniques for quantum control and
  computation},
\newblock \bibinfo{journal}{Rev. Mod. Phys.} \bibinfo{volume}{76}
  (\bibinfo{year}{2005}) \bibinfo{pages}{1037}.
\bibitem[{Ryan et~al.(2008)Ryan, Negrevergne, Laforest, Knill, and
  Laflamme}]{ryan2008liquid}
\bibinfo{author}{C.~Ryan}, \bibinfo{author}{C.~Negrevergne},
  \bibinfo{author}{M.~Laforest}, \bibinfo{author}{E.~Knill},
  \bibinfo{author}{R.~Laflamme},
\newblock \bibinfo{title}{Liquid-state nuclear magnetic resonance as a testbed
  for developing quantum control methods},
\newblock \bibinfo{journal}{Phys. Rev. A.} \bibinfo{volume}{78}
  (\bibinfo{year}{2008}) \bibinfo{pages}{012328}.
\bibitem[{Schanda and Ernst(2016)}]{schanda2016studying}
\bibinfo{author}{P.~Schanda}, \bibinfo{author}{M.~Ernst},
\newblock \bibinfo{title}{Studying dynamics by magic-angle spinning solid-state
  {NMR} spectroscopy: Principles and applications to biomolecules},
\newblock \bibinfo{journal}{Prog. Nucl. Magn. Reson. Spectrosc.}
  \bibinfo{volume}{96} (\bibinfo{year}{2016}) \bibinfo{pages}{1--46}.
\bibitem[{Zobov and Lundin(2014)}]{Zobov2014}
\bibinfo{author}{V.~E. Zobov}, \bibinfo{author}{A.~A. Lundin},
\newblock \bibinfo{title}{Quantum and classical correlations in the solid-state
  {NMR} free induction decay},
\newblock \bibinfo{journal}{Appl. Magn. Reson.} \bibinfo{volume}{45}
  (\bibinfo{year}{2014}) \bibinfo{pages}{1169}.
\bibitem[{Liu et~al.(2012)Liu, Pan, Jiang, Zhao, and Liu}]{Liu2012}
\bibinfo{author}{G.-Q. Liu}, \bibinfo{author}{X.-Y. Pan},
  \bibinfo{author}{Z.-F. Jiang}, \bibinfo{author}{N.~Zhao},
  \bibinfo{author}{R.-B. Liu},
\newblock \bibinfo{title}{Controllable effects of quantum fluctuations on spin
  free-induction decay at room temperature},
\newblock \bibinfo{journal}{Sci. Rep.} \bibinfo{volume}{2}
  (\bibinfo{year}{2012}) \bibinfo{pages}{432}.
\bibitem[{Slichter(1990)}]{Slichter1990}
\bibinfo{author}{C.~P. Slichter}, \bibinfo{title}{Principles of Magnetic
  Resonance}, \bibinfo{publisher}{Springer Berlin Heidelberg},
  \bibinfo{year}{1990}.
\bibitem[{Chen et~al.(2005)Chen, Marble, Colpitts, and Balcom}]{Chen2005}
\bibinfo{author}{Q.~Chen}, \bibinfo{author}{A.~Marble},
  \bibinfo{author}{B.~Colpitts}, \bibinfo{author}{B.~Balcom},
\newblock \bibinfo{title}{The internal magnetic field distribution, and single
  exponential magnetic resonance free induction decay, in rocks},
\newblock \bibinfo{journal}{J. Magn. Reson.} \bibinfo{volume}{175}
  (\bibinfo{year}{2005}) \bibinfo{pages}{300--308}.
\bibitem[{Schenzle et~al.(1980)Schenzle, Wong, and Brewer}]{Schenzle1980}
\bibinfo{author}{A.~Schenzle}, \bibinfo{author}{N.~C. Wong},
  \bibinfo{author}{R.~G. Brewer},
\newblock \bibinfo{title}{Oscillatory free-induction decay},
\newblock \bibinfo{journal}{Phys. Rev. A.} \bibinfo{volume}{21}
  (\bibinfo{year}{1980}) \bibinfo{pages}{887}.
\bibitem[{Kunitomo et~al.(1982)Kunitomo, Endo, Nakanishi, and
  Hashi}]{Kunitomo1982}
\bibinfo{author}{M.~Kunitomo}, \bibinfo{author}{T.~Endo},
  \bibinfo{author}{S.~Nakanishi}, \bibinfo{author}{T.~Hashi},
\newblock \bibinfo{title}{Oscillatory free-induction decay and oscillatory spin
  echoes},
\newblock \bibinfo{journal}{Phys. Rev. A.} \bibinfo{volume}{25}
  (\bibinfo{year}{1982}) \bibinfo{pages}{2235}.
\bibitem[{Boscaino and Gelardi(1987)}]{Boscaino1987}
\bibinfo{author}{R.~Boscaino}, \bibinfo{author}{F.~M. Gelardi},
\newblock \bibinfo{title}{Experimental spectra of oscillatory free-induction
  decay},
\newblock \bibinfo{journal}{Phys. Rev. A.} \bibinfo{volume}{35}
  (\bibinfo{year}{1987}) \bibinfo{pages}{3561}.
\bibitem[{Sabirov and Vinukhov(1982)}]{Sabirov1982}
\bibinfo{author}{R.~K. Sabirov}, \bibinfo{author}{I.~A. Vinukhov},
\newblock \bibinfo{title}{On the nature of the oscillating behaviour of the
  free induction decay},
\newblock \bibinfo{journal}{Phys. Status Solidi B} \bibinfo{volume}{110}
  (\bibinfo{year}{1982}) \bibinfo{pages}{281--284}.
\bibitem[{Cho et~al.(2005)Cho, Ladd, Baugh, Cory, and Ramanathan}]{Cho2005}
\bibinfo{author}{H.~Cho}, \bibinfo{author}{T.~D. Ladd},
  \bibinfo{author}{J.~Baugh}, \bibinfo{author}{D.~G. Cory},
  \bibinfo{author}{C.~Ramanathan},
\newblock \bibinfo{title}{Multispin dynamics of the solid-state nmr free
  induction decay},
\newblock \bibinfo{journal}{Phys. Rev. B.} \bibinfo{volume}{72}
  (\bibinfo{year}{2005}) \bibinfo{pages}{054427}.
\bibitem[{Oron et~al.(2002)Oron, Dudovich, Yelin, and
  Silberberg}]{oron2002quantum}
\bibinfo{author}{D.~Oron}, \bibinfo{author}{N.~Dudovich},
  \bibinfo{author}{D.~Yelin}, \bibinfo{author}{Y.~Silberberg},
\newblock \bibinfo{title}{Quantum control of coherent {anti-Stokes Raman}
  processes},
\newblock \bibinfo{journal}{Phys. Rev. A.} \bibinfo{volume}{65}
  (\bibinfo{year}{2002}) \bibinfo{pages}{043408}.
\bibitem[{Wollenhaupt et~al.(2005)Wollenhaupt, Pr{\"a}kelt, Sarpe-Tudoran,
  Liese, and Baumert}]{wollenhaupt2005quantum}
\bibinfo{author}{M.~Wollenhaupt}, \bibinfo{author}{A.~Pr{\"a}kelt},
  \bibinfo{author}{C.~Sarpe-Tudoran}, \bibinfo{author}{D.~Liese},
  \bibinfo{author}{T.~Baumert},
\newblock \bibinfo{title}{Quantum control and quantum control landscapes using
  intense shaped femtosecond pulses},
\newblock \bibinfo{journal}{J. Mod. Opt.} \bibinfo{volume}{52}
  (\bibinfo{year}{2005}) \bibinfo{pages}{2187--2195}.
\bibitem[{Dudovich et~al.(2004)Dudovich, Oron, and
  Silberberg}]{dudovich2004quantum}
\bibinfo{author}{N.~Dudovich}, \bibinfo{author}{D.~Oron},
  \bibinfo{author}{Y.~Silberberg},
\newblock \bibinfo{title}{Quantum control of the angular momentum distribution
  in multiphoton absorption processes},
\newblock \bibinfo{journal}{Phys. Rev. Lett.} \bibinfo{volume}{92}
  (\bibinfo{year}{2004}) \bibinfo{pages}{103003}.
\bibitem[{Abragam(1961)}]{abragam1961principles}
\bibinfo{author}{A.~Abragam}, \bibinfo{title}{The Principles of Nuclear
  Magnetism}, \bibinfo{publisher}{Oxford University Press},
  \bibinfo{year}{1961}.
\bibitem[{Yang et~al.(2016)Yang, Ma, and Liu}]{yang2016quantum}
\bibinfo{author}{W.~Yang}, \bibinfo{author}{W.-L. Ma}, \bibinfo{author}{R.-B.
  Liu},
\newblock \bibinfo{title}{Quantum many-body theory for electron spin
  decoherence in nanoscale nuclear spin baths},
\newblock \bibinfo{journal}{Rep. Prog. Phys.} \bibinfo{volume}{80}
  (\bibinfo{year}{2016}) \bibinfo{pages}{016001}.
\bibitem[{Messiah(1961)}]{messiah1961quantum}
\bibinfo{author}{A.~Messiah}, \bibinfo{title}{Quantum Mechanics},
  \bibinfo{publisher}{Elsevier Science}, \bibinfo{year}{1961}.
\bibitem[{Dobrovitski et~al.(2008)Dobrovitski, Feiguin, Awschalom, and
  Hanson}]{dobrovitski2008decoherence}
\bibinfo{author}{V.~Dobrovitski}, \bibinfo{author}{A.~Feiguin},
  \bibinfo{author}{D.~Awschalom}, \bibinfo{author}{R.~Hanson},
\newblock \bibinfo{title}{Decoherence dynamics of a single spin versus spin
  ensemble},
\newblock \bibinfo{journal}{Phys. Rev. B} \bibinfo{volume}{77}
  (\bibinfo{year}{2008}) \bibinfo{pages}{245212}.
\bibitem[{Cory et~al.(2000)Cory, Laflamme, Knill, Viola, Havel, Boulant,
  Boutis, Fortunato, Lloyd, Martinez et~al.}]{cory2000nmr}
\bibinfo{author}{D.~G. Cory}, \bibinfo{author}{R.~Laflamme},
  \bibinfo{author}{E.~Knill}, \bibinfo{author}{L.~Viola},
  \bibinfo{author}{T.~Havel}, \bibinfo{author}{N.~Boulant},
  \bibinfo{author}{G.~Boutis}, \bibinfo{author}{E.~Fortunato},
  \bibinfo{author}{S.~Lloyd}, \bibinfo{author}{R.~Martinez}, et~al.,
\newblock \bibinfo{title}{{NMR} based quantum information processing:
  Achievements and prospects},
\newblock \bibinfo{journal}{Fortschritte der Physik: Progress of Physics}
  \bibinfo{volume}{48} (\bibinfo{year}{2000}) \bibinfo{pages}{875--907}.
\bibitem[{Jones(2010)}]{jones2010quantum}
\bibinfo{author}{J.~A. Jones},
\newblock \bibinfo{title}{Quantum computing with {NMR}},
\newblock \bibinfo{journal}{arXiv preprint arXiv:1011.1382}
  (\bibinfo{year}{2010}).
\bibitem[{Cory et~al.(1998)Cory, Price, and Havel}]{cory1998nuclear}
\bibinfo{author}{D.~G. Cory}, \bibinfo{author}{M.~D. Price},
  \bibinfo{author}{T.~F. Havel},
\newblock \bibinfo{title}{Nuclear magnetic resonance spectroscopy: An
  experimentally accessible paradigm for quantum computing},
\newblock \bibinfo{journal}{Physica D} \bibinfo{volume}{120}
  (\bibinfo{year}{1998}) \bibinfo{pages}{82--101}.
\bibitem[{Luo et~al.(2016)Luo, Lei, Li, Nie, Li, Peng, and Du}]{Luo16}
\bibinfo{author}{Z.~Luo}, \bibinfo{author}{C.~Lei}, \bibinfo{author}{J.~Li},
  \bibinfo{author}{X.~Nie}, \bibinfo{author}{Z.~Li}, \bibinfo{author}{X.~Peng},
  \bibinfo{author}{J.~Du},
\newblock \bibinfo{title}{Experimental observation of topological transitions
  in interacting multispin systems},
\newblock \bibinfo{journal}{Phys. Rev. A.} \bibinfo{volume}{93}
  (\bibinfo{year}{2016}) \bibinfo{pages}{052116}.
\bibitem[{Knill et~al.(1998)Knill, Chuang, and Laflamme}]{Knill97}
\bibinfo{author}{E.~Knill}, \bibinfo{author}{I.~Chuang},
  \bibinfo{author}{R.~Laflamme},
\newblock \bibinfo{title}{Effective pure states for bulk quantum computation},
\newblock \bibinfo{journal}{Phys. Rev. A.} \bibinfo{volume}{57}
  (\bibinfo{year}{1998}) \bibinfo{pages}{3348--3363}.

\end{thebibliography}

\bio{}
\endbio

\endbio

\end{document}